\documentclass[11pt,a4paper]{article}
\pdfoutput=1

\usepackage{jheppub}

\usepackage{graphicx, epsfig} %include figure files
\usepackage{amsmath,amssymb,amsfonts,dsfont,mathrsfs,amsthm,mathtools}
\usepackage{bm} %include bold math: \bm{} creates bold letters in math mode
\usepackage{color}
\usepackage{float}
\usepackage{xcolor}
\usepackage{hyperref}
\usepackage{siunitx}
\usepackage[normalem]{ulem}
\hypersetup{colorlinks=true,urlcolor=blue,linkcolor=magenta,citecolor=blue,filecolor=blue}
\usepackage[normalem]{ulem}
\usepackage{array}
\usepackage{hyperref}
\usepackage{booktabs}
\usepackage{blindtext}
\usepackage{amsfonts}
\usepackage{tensor}
\usepackage{comment}
\usepackage{bbding}

\allowdisplaybreaks

\usepackage{float}
\usepackage[scr=rsfs]{mathalfa}
\usepackage{bigints}

\newcommand*{\tr}{\operatorname{tr}}

\newcommand{\diff}[1]{\text{d}#1}
\newcommand{\dd}{\text{d}}

\newcommand{\Lie}{\mathcal{L}}

\newcommand{\Lag}{\mathscr{L}}

\newcommand{\sqrtgbdy}{\sqrt{{\scriptstyle |}g_{(0)} {\scriptstyle |} }}

\begin{document}

%%%%%%%
\title{Extrinsic Renormalization of New Massive Gravity}

\author[a]{Crist\'obal Corral,}
\author[b]{Rodrigo Olea,}
\author[c]{Leonardo Sanhueza}

\affiliation[a]{Departamento de Ciencias, Facultad de Artes Liberales, Universidad Adolfo Ibáñez, \\ Avenida Padre Hurtado 750, 2562340, Viña del Mar, Chile}

\affiliation[b]{Instituto de F\'isica, Pontificia Universidad Cat\'olica de Valpara\'iso, Casilla 4059, Valpara\'iso, Chile}

\affiliation[c]{Departamento de F\'isica, Universidad de Concepci\'on, Casilla, 160-C, Concepci\'on, Chile}

\vspace{0.3cm}

\emailAdd{cristobal.corral@uai.cl}
\emailAdd{rodrigo\_olea\_a@yahoo.co.uk}
\emailAdd{lsanhueza@udec.cl}

\abstract{
The renormalization of asymptotically AdS spacetimes in New Massive Gravity is studied using boundary terms constructed with the extrinsic curvature. For generic values of the theory couplings, a single extrinsic boundary term produces a consistent holographic description, as the variation of the total action is finite and expressible in terms of the holographic source.  This prescription applies to constant-curvature geometries and non-Einstein solutions with Brown-Henneaux asymptotic conditions. A suitable asymptotic expansion shows the agreement of the resulting holographic stress tensor with the one obtained from the auxiliary-field formulation.

At the degenerate point, where the two maximally symmetric vacua coalesce, the Fefferman-Graham expansion admits an additional mode associated with relaxed AdS boundary conditions. A quadratic extrinsic counterterm removes the new divergences and leads naturally to a holographic picture with two independent sources. The corresponding responses obey a modified holographic Ward identity and determine finite conserved charges.

Finally, the proper use of the Noether-Wald formalism reproduces the correct charges of the BTZ black hole, non-Einstein AdS waves, and the rotating hairy black hole. 
}
%%%%%%

\maketitle

\section{Introduction}
Three-dimensional gravity is a theoretical laboratory for testing, in a tractable manner, different classical and quantum gravitational features of gravitation. The Ba\~nados-Teitelboim-Zanelli (BTZ) black hole is, perhaps, the best example of nontrivial properties, encoded in a yet simple geometry~\cite{Banados:1992wn,Banados:1992gq}. It resembles several properties of the Kerr metric while being locally equivalent to three-dimensional anti-de Sitter space (AdS). Indeed, it provides a controlled setup in which one can obtain its microscopic entropy without supersymmetry~\cite{Strominger:1997eq}. Furthermore, the one-loop determinant for the graviton and higher-spin fields on AdS$_3$ has been studied using different methods and boundary conditions~\cite{Maloney:2007ud,Yin:2007gv,Giombi:2008vd,David:2009xg,Datta:2011za,Castro:2017mfj,Acosta:2021oqt,Acito:2025hka}, allowing one to extract the one-loop corrections to the semiclassical entropy of the (Euclidean) BTZ black hole. In point of fact, asymptotic symmetries in AdS$_3$ with suitable boundary conditions gave a first glimpse of how a nontrivial Virasoro central charge arises in the algebra of canonical generators~\cite{Brown:1986nw}.

The simplicity of three-dimensional Einstein gravity largely stems from its lack of local degrees of freedom. However, higher-curvature corrections, which arise naturally from an effective field theory viewpoint, change this picture drastically. Along this line, there exists an interesting parity-even extension of three-dimensional General Relativity --- known as New Massive Gravity (NMG)~\cite{Bergshoeff:2009hq} --- whose higher-curvature terms induce two massive spin-2 fields, endowing the theory with local degrees of freedom. The BTZ black hole still solves the NMG field equations, since the theory admits constant-curvature spaces as solutions. Nonetheless, NMG also accommodates non-Einstein AdS waves~\cite{Ayon-Beato:2009cgh}, non-constant curvature black hole solutions with weakened AdS boundary conditions~\cite{Oliva:2009ip,Giribet:2009qz,Bergshoeff:2009aq}, asymptotically Lifshitz black holes~\cite{Ayon-Beato:2009rgu}, solitons~\cite{Perez:2011qp}, and warped AdS black holes~\cite{Clement:2009gq} which serve as a lower-dimensional toy model to account for its microscopic entropy~\cite{Donnay:2015iia}. The richness of NMG provides interesting examples for studying holography beyond the AdS/CFT correspondence~\cite{Maldacena:1997re,Witten:1998qj,Gubser:1998bc}, which are relevant for Condensed Matter systems and for understanding the microscopic origin of black hole thermodynamics.

In the path-integral formulation of quantum gravity, a well-posed variational principle for the boundary fields is mandatory. This requirement demands that the action be extremized by solutions to the equations of motion under Dirichlet boundary conditions. In Einstein gravity, this is achieved by adding the Gibbons-Hawking-York~\cite{York:1972sj,Gibbons:1976ue} term, which defines the canonical momentum at the boundary. However, for asymptotically AdS spacetimes, a Dirichlet condition on the metric is ill defined due to its divergent behavior at the conformal boundary~\cite{Papadimitriou:2005ii}. Therefore, the variational principle needs to be defined in terms of the metric at the conformal boundary, instead. This simple observation makes room to the addition of extrinsic counterterms, as the variation of the total action may be still holographic, i.e., finite and expressed as variations of the holographic source~\cite{Olea:2006vd}.

Indeed, in even bulk dimensions, extrinsic boundary terms associated to topological invariants of the Euler class act as counterterms~\cite{Aros:1999id,Olea:2005gb,Miskovic:2009bm,Anastasiou:2020zwc}.  In odd dimensions, the use transgression forms produces boundary contributions which are, once again, extrinsic~\cite{Olea:2006vd}. In general, this method matches the results of holographic renormalization for Weyl-flat boundaries~\cite{Anastasiou:2020zwc}. In higher even dimensions, and for generic boundary geometries, this prescription is superseded by conformal structures in the bulk, which matches the results of standard holographic techniques~\cite{Anastasiou:2020mik,Anastasiou:2023oro,Anastasiou:2025usa,Anastasiou:2026jrt}.  In the presence of higher-curvature terms, renormalization with extrinsic counterterms in asymptotically AdS spaces has been studied in Refs.~\cite{Giribet:2018hck,Araya:2021atx,Miskovic:2022mqv,Miskovic:2023hzc}. This is somewhat natural, as a higher-derivative gravity theory does not lend itself to a Dirichlet problem for the metric, making renormalization difficult with intrinsic counterterms.

The parameter space in NMG contains points where either the central charges vanish or the two maximally symmetric vacua coalesce. These are referred to as the critical and degenerate points, respectively. In both cases, additional modes in the Fefferman-Graham expansion are turned on: in the former, log modes might appear, while in the latter, a linear mode arises, both leading to weakened versions of the Brown-Henneaux boundary conditions~\cite{Brown:1986nw}. In particular, it is  at the degenerate point where the hairy black hole with non-constant curvature exists in three dimensions~\cite{Oliva:2009ip,Giribet:2009qz,Bergshoeff:2009aq}. 

Renormalization of NMG out of these special points has been studied in Ref.~\cite{Hohm:2010jc} by using the auxiliary-field method. At the degenerate point, however, the situation is more subtle: the field equations yield an additional holographic source at the conformal boundary. This was discussed in Refs.~\cite{Alishahiha:2010bw,Giribet:2010ed,Kwon:2011jz,Cunliff:2013en} by using the auxiliary variables formalism developed in the original Ref.~\cite{Bergshoeff:2009aq} for generic quadratic gravity theories in three dimensions. In Ref.~\cite{Araya:2021atx,Miskovic:2022mqv}, on the other hand, renormalization of higher-curvature gravity in three dimensions was studied for general values of the coupling constants such that the linear term in the Fefferman-Graham expansion vanishes. 

In this work, extrinsic boundary counterterms are constructed to renormalize NMG under different boundary conditions, including weakened AdS falloff triggered by the vacuum degeneracy. The resulting prescription defines a holographic variational principle for the sources and matches holographic responses found by different methods. The analysis first revisits the extrinsic renormalization of three-dimensional Einstein gravity. At generic couplings, it determines the boundary term that renders the action and its variations finite for asymptotically AdS solutions of NMG. An explicit evaluation of non-Einstein AdS waves confirms the finiteness of the associated conserved charges~\cite{Ayon-Beato:2009cgh}. The degenerate branch with relaxed boundary conditions is subsequently analyzed by canceling the divergences generated by the linear Fefferman--Graham coefficient and deriving the holographic Ward identities associated with boundary diffeomorphisms. As an example, the conserved charges of the hairy black hole\cite{Oliva:2009ip,Giribet:2009qz} are computed. They are shown to agree with different methods in the literature. 

The manuscript is organized as follows. Section~\ref{sec:3D-Einstein} reviews the role of extrinsic boundary terms in three-dimensional Einstein-AdS gravity. Section ~\ref{sec:NMG} introduces NMG and discusses its constant-curvature vacua, together with the degenerate and critical points of its parameter space. Section~\ref{sec:asymptotics} develops the asymptotic analysis of the field equations and constructs the boundary counterterms required to renormalize the action and its variations for the different asymptotic sectors, including the degenerate branch with relaxed AdS boundary conditions. The resulting holographic description is also compared with the one obtained from the auxiliary-field formulation. Section~\ref{sec:NW} derives the corresponding Noether-Wald charges, while Sec.~\ref{Sec.BHandConservedCharges} evaluates them for representative solutions, including the BTZ black hole, nonconstant-curvature AdS waves, and the hairy black hole. Finally, Sec.~\ref{sec:conclusions} contains concluding remarks.

\section{Boundary terms in Einstein-AdS\texorpdfstring{$_3$}{3} gravity\label{sec:3D-Einstein}}
Einstein gravity in three dimensions has no propagating degrees of freedom, irrespective of the value of the cosmological constant. However, solutions in the bulk with vanishing field strength (for dS, Poincaré, or AdS groups) do not imply a trivial theory. Indeed, for a given set of asymptotic conditions, three-dimensional gravity may induce a nontrivial dynamics at the boundary. Therefore, properly identifying boundary terms and boundary conditions is key to generating modes of physical interest in the asymptotic region.

In order to describe the gravitational dynamics in asymptotically AdS spacetimes, one may consider a foliation of the spacetime in terms of Gauss-normal coordinates. In this frame, the induced metric $h_{ij}$ is defined by the radial evolution of hypersurfaces as
\begin{align}\label{GNcoord}
    \dd s^2 = N^2(z)\dd z^2 + h_{ij}(z,x)\dd x^i \dd x^j\,,
\end{align}
where $z$ and $\{x^i\}$ denote radial and codimension-1 boundary coordinates, respectively. For a fixed value of $z$, the intrinsic geometry defines the one of a boundary  $\partial\mathcal{M}$.

The corresponding action is given by a bulk term and a boundary contribution, i.e.,
\begin{align}\label{EAdSaction}
I[g_{\mu\nu}] = \kappa \int_{\mathcal{M}}\dd^3x\sqrt{|g|}\left(R-2\lambda \right) + \int_{\partial\mathcal{M}}\dd^2x\sqrt{|h|}\;\mathcal{B}\,,   
\end{align}
where $g=\det g_{\mu\nu}$ is the metric determinant, $\kappa=(16 \pi G)^{-1}$ is the gravitational coupling with $G$ being the Newton constant, $\lambda=-L^{-2}$ is the cosmological constant (with $L$ being the AdS radius),  and $h=\det h_{ij}$ is the determinant of the boundary metric. The role of $\mathcal{B}$ is to define a well-posed variational principle for Dirichlet boundary conditions on the metric. This stems from the fact that the bulk action, evaluated in the coordinate system~\eqref{GNcoord}, contains a linear term in the second derivative of $h_{ij}$ in the normal direction $z$. As a consequence, a well-defined action principle with Dirichlet boundary conditions for the metric $h_{ij}$ requires the addition of the Gibbons-Hawking-York (GHY) term~\cite{York:1972sj,Gibbons:1976ue}. Furthermore, in asymptotically AdS gravity, the renormalization of both the action and its variation requires the addition of local counterterms as surface terms~\cite{Balasubramanian:1999re,Emparan:1999pm,deHaro:2000vlm}, that is, $\mathcal{B} = \mathcal{B}_{\rm GHY} + \mathcal{B}_{\rm ct}$, where
\begin{align} \label{GHY}
 \mathcal{B}_{\rm GHY}=-\frac{1}{8\pi G}\,K \qquad \mbox{and}\qquad  \mathcal{B}_{\rm ct} = \frac{1}{8\pi GL}\,, 
\end{align}
with $K=h^{ij}K_{ij}$ being the trace of the extrinsic curvature defined in the coordinate frame~\eqref{GNcoord} as $K_{ij}=-\frac{1}{2N}\partial_z h_{ij}$. The second term, $\mathcal{B}_{\rm ct}$, removes infrared divergences and allows for the computation of holographic correlators~\cite{Balasubramanian:1999re,Emparan:1999pm,deHaro:2000vlm,Bianchi:2001kw}.

Arbitrary variations of the action~\eqref{EAdSaction} with respect to the metric yields 
\begin{equation}
   \delta I_{\rm ren}=\kappa\int_{\mathcal{M}} \dd^{3}x\sqrt{|g|}\left(G_{\mu\nu}+\Lambda g_{\mu\nu}\right)\delta g^{\mu\nu}+\kappa\int_{\partial\mathcal{M}}\dd^{2}x\sqrt{|h|}\left(K_{j}^{i}-\delta_{j}^{i}K + \frac{1}{L}\delta^i_j \right)\left(h^{-1}\delta h \right)_{i}^{j}\,,
\end{equation}
where $G_{\mu\nu}=R_{\mu\nu}-\tfrac{1}{2}g_{\mu\nu}R$ is the Einstein tensor and the boundary terms in Eq.~\eqref{GHY} have been used. 

It may be rendered manifest that, in AAdS spaces, the canonical momentum (given by the particular combination of the extrinsic curvature and its trace) contains infinities in the limit $z\to0$, which can be properly renormalized in the FG frame by the counterterm $\mathcal{B}_{\rm ct}$~\cite{deHaro:2000vlm,Bianchi:2001kw}. This feature can be readily analyzed in the Fefferman-Graham (FG) gauge, that is,
\begin{subequations}\label{FGexp}
\begin{align}
    \dd s^2 = \frac{L^2}{z^2}\left(\dd z^2 + \bar{g}_{ij}(z,x)\dd x^i \dd x^j \right)\,, \label{eq.FGG}
\end{align}
where the metric $\bar{g}_{ij}(z,x)$, defined in terms of the induced metric as $h_{ij}=\tfrac{L^2}{z^2}\bar{g}_{ij}$, admits a power series expansion near the conformal boundary at $z\to0$, of the form
\begin{align} \label{eq.expansion}
\bar{g}_{ij}(z,x) = g_{(0)ij}(x) + \frac{z}{L}g_{(1)ij}  + \frac{z^2}{L^2}\left(g_{(2)ij} + a_{(2)ij}\log z \right) + \mathcal{O}(z^3)\,.
\end{align}
\end{subequations}
Using the definition of the extrinsic curvature given above, one could read the asymptotic behavior of the latter, resulting in 
\begin{align}\label{K-FG}
K^i_j=  \frac{1}{L}\delta^i_j -  \frac{z}{2L^2}g^i_{(1)j} +  \frac{z^2}{L^3} \left( \frac{1}{2} g_{(1)k}^i  g_{(1)j}^k -  g_{(2)j}^i - \frac12 a_{(2)j}^i-a_{(2)j}^i \log z  \right) +\mathcal{O}(z^3)\,.   
\end{align} 
In pure Einstein-AdS$_3$ gravity, for Brown-Henneaux boundary conditions~\cite{Brown:1986nw},  both modes $g_{(1)ij}$ and $a_{(2)ij}$ vanish, what is implied by the asymptotic resolution of the field equations. However, this is not necessarily true when one includes higher-curvature corrections, as one may see next. Additionally, the $zz$ component of the Einstein equations implies that
\begin{align} \label{trg2condition}
\tr g_{(2)}=-\frac{L^2}{2}\,\mathcal{R}_{(0)}  \,,  
\end{align}
where $\mathcal{R}_{(0)}$ is the Ricci scalar associated to the boundary metric $g_{(0)ij}$. Then, using the asymptotic expansion of the determinant of the induced metric
\begin{equation} \label{sqrtdethexp}
    \sqrt{|h|} = \frac{L^2}{z^2} \sqrtgbdy \left[ 1 + \frac{z}{2 L} \tr g_{(1)} + \frac{z^2}{2 L^2} \left( \tr g_{(2)} + \frac{1}{4} (\tr g_{(1)})^2 - \frac{1}{2}\tr g_{(1)}^2   \right)   + \mathcal{O}(z^3) \right]\,,
\end{equation}
and imposing the condition~\eqref{trg2condition} along with the fact that $g_{(1)ij}=0$, the on-shell variation of the renormalized Einstein-AdS$_3$ action can be written as
\begin{align}\label{FG-GHY}
    \delta I_{\rm{ren}}  &= \frac{\kappa}{L}\int_{\partial \mathcal{M}}\diff{^2}x\sqrtgbdy\;\delta g_{(0)ij}\left(g_{(2)}^{ij} -g_{(0)}^{ij} \tr g_{(2)} \right)  := \frac{1}{2}\int_{\partial \mathcal{M}}\diff{^2}x\sqrtgbdy\;\delta g_{(0)ij}\,\tau^{ij}_{\rm (E)} \,.
\end{align}
The response to the holographic source $\tau^{ij}_{(E)}$ is the stress tensor in the dual CFT, as dictated within the context of AdS/CFT correspondence~\cite{Maldacena:1997re,Gubser:1998bc,Witten:1998qj}. 

Any modification to the factor of $\mathcal{B}_{\rm GHY}$ in Eq.~\eqref{GHY} would spoil a variational principle based on a Dirichlet condition on the induced metric $h_{ij}$. Then, at first glance, it is puzzling that half of the GHY term
\begin{align}\label{GHY/2}
\mathcal{B}= - \frac{1}{16\pi G}\,K    \,,
\end{align}
renders the Euclidean action finite for AAdS solutions~\cite{Banados:1998ys}. This fact can be explained by resorting to a variational principle based on Dirichlet boundary conditions for $g_{(0)}$ instead, which also renormalizes the action and conserved charges of asymptotically AdS$_3$ spacetimes in Einstein gravity~\cite{Miskovic:2006tm,Olea:2006vd}. This can be checked by noticing that the on-shell variation of the action with one-half of the GHY term, defined as $\tilde{I}_{\rm ren}$, yields 
\begin{equation} \label{halfGHY}
   \delta \tilde{I}_{\rm ren}=\kappa\int_{\partial\mathcal{M}}\diff{}^{2}x\sqrt{|h|}\left[\left(K_{j}^{i}-\frac{1}{2}\delta_{j}^{i}K\right)\left(h^{-1}\delta h\right)_{i}^{j} + \delta^i_j \delta K^j_i   \right]\,,
\end{equation}
After performing the FG expansion and using the solutions obtained by solving the field equations asymptotically order by order, one finds that this equation reduces to 
\begin{align} \label{deltaIEtilde}
    \delta \tilde{I}_{\rm ren} &=-   \frac{\kappa}{L} \int_{\partial \mathcal{M}} \diff{}^2x \left[ \sqrtgbdy\; \delta g_{(0)ij} \left(g_{(2)}^{ij} -g_{(0)}^{ij} \tr g_{(2)} \right)+ \frac{L^2}{2}\delta \left( \sqrtgbdy\;  \mathcal{R}_{(0)} \right)+ \mathcal{O}(z)\right] \,.
\end{align}
Notice that the third term corresponds to the variation of the boundary Euler density, which is topological. Thus, the above result implies that the boundary terms Eqs.~\eqref{GHY} and~\eqref{GHY/2}, give rise to the same dynamics at the conformal boundary. However, considering a Dirichlet problem for the holographic source gives room for an extrinsic renormalization of the gravity action, while reproducing the same holographic output. Additionally, boundary diffeomorphism invariance of $g_{(0)ij}$ along the flow generated by the vector field $\xi=\xi^i\partial_i$ yields the holographic Ward identity $\mathcal{D}^{(0)}_i\tau^{ij}_{\rm (E)}=0$. Conserved charges, on the other hand, can be obtained by computing
\begin{align}\label{chargetauijNMG}
    Q[\xi] = \int_\Sigma\diff{x}\sqrt{|\gamma|}\,u^i\,\tau_{ij}^{\rm (E)}\,\xi^j\,,
\end{align}
where $\gamma$ is the metric determinant of the codimension-2 hypersurface $\Sigma$ with timelike unit normal $u=u^i\partial_i$. This paper aims at showing that this prescription extends to New Massive Gravity in three dimensions.

\section{New Massive Gravity\label{sec:NMG}}

In three dimensions, New Massive Gravity (NMG) is a parity-preserving higher-curvature theory that, unlike Einstein-AdS gravity, propagates dynamical degrees of freedom~\cite{Bergshoeff:2009hq,Bergshoeff:2009aq}. Its linearized theory around a maximally symmetric vacuum is equivalent to the unitary Fierz-Pauli action for a spin-2 field. Its bulk dynamics is governed by the action
\begin{align} 
    I_{\rm NMG} &= \kappa  \int_\mathcal{M} \diff{}^3x \sqrt{|g|}\left(\sigma_1 R- 2\lambda - \frac{\sigma_2}{\mu^2} \left[ R_{\mu \nu} R^{\mu \nu} - \frac{3}{8} R^2\right] \right) :=   \int_\mathcal{M} \diff{}^3x \sqrt{|g|}\,\Lag_{\rm NMG}\label{INMG} \,,
\end{align}
where $\mu$ sets the mass scale of the higher curvature sector which, at the linearized level, corresponds to the mass of a unitary Fierz-Pauli spin-2 field. The parameters $\sigma_{1,2} =\pm1$ are introduced to accommodate the sign conventions commonly found in the literature. 

As shown in Ref.~\cite{Tekin:2015rha}, the Cotton and Schouten tensors provide a convenient basis to write down the NMG action and field equations. This stems from the fact that, as the Weyl tensor is identically zero in three dimensions, the Riemann tensor is fully determined by the Schouten tensor, that is, $R^{\mu\nu}_{\lambda\rho}=4\delta^{[\mu}_{[\lambda}S^{\nu]}_{\rho]}$ where
\begin{equation} \label{Schouten}
    S^\mu_\nu = R^\mu_\nu - \frac{1}{4} \delta^\mu_\nu R\,,
\end{equation}
whose trace is $S\equiv S^\mu_\mu = \tfrac{1}{4} R$. Thus, one can rewrite the action~\eqref{INMG} in a more convenient form, using the fact that the curvature-squared combination can be expressed as 
\begin{equation}
     R_{\mu \nu} R^{\mu \nu} - \frac{3}{8} R^2   = -\, \delta^{\mu \nu}_{\lambda \rho} S^\lambda_\mu S^\rho_\nu \,, 
\end{equation}
where the generalized Kronecker delta and its relevant properties are collected in Appendix~\ref{sec:gen-kdelta}. 
Consequently, the NMG Lagrangian written solely in terms of the Schouten tensor is 
\begin{equation}\label{LagNMG}
    \Lag_{\rm NMG} = \kappa\left(4\sigma_1 S -2\lambda + \frac{\sigma_2}{\mu ^2} \delta^{\mu \nu}_{\lambda \rho} S^\lambda_\mu S^\rho_\nu\right)  \,.
\end{equation}

At linear order, the field equations of NMG coincide with those of unitary Fierz-Pauli theory for a massive spin-2 field~\cite{Bergshoeff:2009hq}. At nonlinear order, an arbitrary variation of the action~\eqref{INMG} produces
\begin{equation}\label{deltaINMG}
    \delta I_{\rm NMG} = \int_\mathcal{M} \dd^3 x \sqrt{|g|}\, \mathcal{E}_{\mu \nu} \delta g^{\mu \nu} + \int_\mathcal{M} \dd^3 x \sqrt{|g|} \nabla_\mu \Theta^\mu(g,\delta g) \,.
\end{equation}
Here, $\mathcal{E}_{\mu \nu}$ denotes the field equations, and $\Theta^\mu$ is the boundary term arising from integration by parts. In general higher-curvature theories, these quantities can be expressed as~\cite{Padmanabhan:2011ex}
\begin{align}
    \mathcal{E}_{\mu \nu} &= E_\mu{}^{\lambda \rho \sigma} R_{\nu \lambda \rho \sigma}-\frac{1}{2} g_{\mu \nu} \Lag_{\rm NMG} - 2\nabla^\lambda \nabla^\rho E_{\mu \lambda \rho \nu}\,,   \label{Emunu} \\ 
     \Theta^\mu  &= 2 \delta \Gamma^{\lambda}_{\nu \rho} E_{\lambda}^{\ \rho \mu \nu} - 2 \delta g_{\nu \sigma} \nabla_{\rho} E^{\nu \mu \rho \sigma}\,,  \label{Thetamu}
\end{align}
where $E_{\mu\nu\lambda\rho}$ denotes the functional derivative of the Lagrangian with respect to the Riemann tensor. In the case of the NMG Lagrangian as written in Eq.~\eqref{LagNMG}, one finds that the $E$-tensor is given by
\begin{equation}\label{Etensor}
    E^{\mu \nu}_{\lambda \rho} = \frac{\partial \mathcal{L}_{\rm NMG}}{\partial R^{\mu \nu}_{\lambda \rho}}= \frac{\kappa\sigma_1}{2} \delta^{\lambda \rho}_{\mu \nu} - \frac{2\kappa \sigma_2}{\mu^2} \left( \delta^{[\lambda}_{[\mu} S^{\rho]}_{\nu]} - \frac{1}{4} \delta^{\lambda \rho}_{\mu \nu} S\right) \,,
\end{equation}
while its divergence yields 
\begin{equation} \label{nablaEtensor}
    \nabla^\rho E^{\mu\nu}_{\lambda\rho} = - \frac{\kappa\sigma_2}{2\mu^2}\,C_{ \lambda}^{\ \mu\nu}\,,
\end{equation}
where $C_{\lambda \mu \nu} = 2 \nabla_{[\nu} S_{\mu] \lambda}$ is the Cotton tensor. Then, replacing these results in Eqs.~\eqref{Emunu} and~\eqref{Thetamu}, one finds that the field equations and boundary terms are 
\begin{align}\label{eom2}
    \kappa^{-1}\,\mathcal{E}^{\mu}_\nu&=\sigma_1\, G^\mu_\nu+\lambda\,\delta_{\nu}^{\mu} + \frac{\sigma_2}{2\mu^{2}}\left(\delta_{\nu\lambda\rho}^{\mu\alpha\beta}S_{\,\alpha}^{\lambda}S_{\,\beta}^{\rho}+2\,\nabla_{\lambda}\tensor{C}{_{\nu}^{\lambda\mu}}\right)\,, \\
    \Theta^\mu &= 2\delta \Gamma^\sigma_{\nu \rho} g^{\rho \alpha} \left[\frac{\kappa\sigma_1}{2} \delta_{\sigma \alpha}^{\mu \nu} - \frac{2\kappa \sigma_2}{\mu^2} \left( \delta^{[\mu}_{[\sigma} S^{\nu]}_{\alpha]} - \frac{1}{4} \delta_{\sigma\alpha}^{\mu \nu} S\right)\right]-  \frac{\kappa\sigma_2}{\mu^2} (g^{-1}\delta g)^\lambda_\nu C_\lambda{}^{ \nu \mu} \, , \label{Thetamu2}
\end{align}
with the shorthand notation $(g^{-1}\delta g)^\lambda_\nu = g^{\lambda\sigma}\delta g_{\sigma\nu}$. In terms of the Cotton and Schouten tensors, the $\mathcal{K}_{\mu\nu}$ tensor of Ref.~\cite{Bergshoeff:2009hq,Tekin:2015rha} can be expressed conveniently as
\begin{align}\notag
    \mathcal{K}^\mu_\nu &= 2 \Box R^\mu_\nu - \frac{1}{2} \nabla^\mu \nabla_\nu R - \frac{1}{2} \delta^\mu_\nu \Box R + 4 R^{\mu \lambda}_{ \nu \rho } R^{ \rho}_\lambda -\frac{3}{2} R R^\mu_{\nu} - \delta^\mu_\nu\left( R_{\lambda \rho} R^{\lambda \rho} - \frac{3}{8} R^2\right) \\
    &= - \left(\delta_{\nu\lambda\rho}^{\mu\alpha\beta}S_{\,\alpha}^{\lambda}S_{\,\beta}^{\rho}+2\nabla_{\lambda}\tensor{C}{_{\nu}^{\lambda\mu}}\right)\,.
\end{align}
Thus, the fourth-derivative terms in the field equations of NMG are encoded in the last term of Eq.~\eqref{eom2}. Despite this fact, for $\sigma_1=-1$, the theory propagates a unitary massive graviton mode around a maximally symmetric background~\cite{Bergshoeff:2009aq}. Additionally, the trace of the field equations~\eqref{eom2} gives 
\begin{align}\label{treom2}
   2\sigma_1 S =  \frac{\sigma_2}{2\mu^2}\delta^{\mu\nu}_{\lambda\rho}S^\lambda_\mu S^\rho_\nu + 3\lambda  \,,
\end{align}
which is second order, even though the NMG field equations are higher order. This property avoids higher-order derivatives of scalar modes, which would have otherwise contributed to the field equations if present~\cite{Bergshoeff:2009hq,Bergshoeff:2009aq}. This property has been used to construct higher-dimensional analogs of NMG, which also produce second-order field equations when evaluated on static and spherically-symmetric spacetimes (see~\cite{Oliva:2010eb,Myers:2010ru,Myers:2010jv} and references thereof).

\subsection{Constant-curvature spaces\label{sec:constant-curvature-spaces}}

It is well-known that the equations of motion~\eqref{eom2} admit constant curvature spaces as solutions, say $R^{\mu \nu}_{\lambda \rho} = \Lambda \,\delta^{\mu \nu}_{\lambda \rho} $, where $\Lambda$ is an effective cosmological constant satisfying
\begin{align}\label{Lambdaeffconstraint}
    \mathcal{E}_0:=\lambda - \sigma_1\Lambda + \frac{\sigma_2\Lambda^2}{4\mu^2}=0\,.
\end{align}
In such case, the Schouten tensor becomes $S_{\mu\nu}=\tfrac{\Lambda}{2}g_{\mu\nu}$ and the Cotton tensor vanishes. Thus, the bare and effective cosmological constants should be carefully distinguished. The latter determines the curvature radius of a given vacuum, while the former appears directly in the action. The two roots of Eq.~\eqref{Lambdaeffconstraint} are 
\begin{equation}\label{Lambdapm}
    \Lambda_{\pm} = 2\sigma_2 \mu^2 \left(\sigma_1 \pm \sqrt{\Delta}  \right) \,, \quad \mbox{where} \quad  \Delta := 1 - \frac{\sigma_2 \lambda}{\mu^2}\,.
\end{equation}

If $\Delta>0$, then the theory admits two distinct maximally symmetric vacua. If $\Delta=0$, these vacua coalesce, while for $\Delta <0$ there is no real maximally symmetric vacuum. From here on, the discussion focuses on the negative cosmological constant case. 
Accordingly, the effective AdS radius is defined by $\Lambda=-\ell^{-2}$, with $\Lambda=\Lambda_\pm$ as given in Eq.~\eqref{Lambdapm}.  

\subsection{Degenerate point}

The condition $\Delta =0$ defines the degenerate point of NMG. The latter fixes $\lambda$ in terms of $\mu^2$ and the effective cosmological constants satisfy $\Lambda_+ = \Lambda_- = 2\sigma_{1} \sigma_2 \mu^2$. It is convenient to define the dimensionless parameter
\begin{align}
    \varpi_{\pm} := \sigma_1 \pm \frac{\sigma_2}{2 \mu^2 \ell^2} \,.
\end{align}
Thus, the degeneracy condition can be equivalently defined as
\begin{equation} \label{degcondition}
    \varpi_{+}  =0 \,. 
\end{equation}
This combination multiplies the next-to-leading order terms in the asymptotic expansion of the field equations that would generically impose the condition $g_{(1)ij}=0$ when $\Delta\neq0$. If Eq.~\eqref{degcondition} is met, then it will have direct consequences on the asymptotic dynamics~\cite{Alishahiha:2010bw,Giribet:2010ed,Kwon:2011jz,Cunliff:2013en}. For instance, the rank of the asymptotic field equations is reduced, allowing the FG expansion to include a nontrivial $g_{(1)ij}$. This, in turn, produces a weaker fall-off than the standard Brown-Henneaux behavior, giving rise to a new holographic source at the conformal boundary. Indeed, the degenerate point is precisely where the static and rotating hairy black holes exist~\cite{Oliva:2009ip,Giribet:2009qz}, which are the main example of  non-Einstein configurations used below.

\subsection{Critical point}

 A different special point in the parameter space is determined by the vanishing of an effective coupling $\varpi=0$, which appears as a multiplicative factor in the conserved charges and the central charge of the asymptotic symmetry algebra. In fact, for Brown-Henneaux boundary conditions~\cite{Brown:1986nw}, both the left- and right-moving copies of the Virasoro algebra that generate the asymptotic algebra of diffeomorphisms acquire the same central charge~\cite{Liu:2009pha,Liu:2009bk,Kraus:2006wn}
\begin{equation}
    c_{L} = c_R = \frac{3 \ell}{2G} \left( \sigma_1 - \frac{\sigma_2}{2 \mu^2 \ell^2} \right) = \frac{3\ell}{2G}\varpi_-\,.
\end{equation} 
Furthermore, as shown in Ref.~\cite{Clement:2009gq}, the Wald charges~\cite{Wald:1993nt,Iyer:1994ys,Wald:1999wa}  for the BTZ black hole also vanish at this point. Charges are nonvanishing only if $\log$ terms are switched on in the metric~\cite{Grumiller:2008es,Grumiller:2008qz,Henneaux:2009pw,Maloney:2009ck,Liu:2009kc}. In that case, NMG admits a black hole solution with logarithmic falloff at the critical point~\cite{Clement:2009ka}, whose properties have been studied in Refs.~\cite{Hohm:2010jc,Giribet:2010ed}. In the present study, however, logarithmic terms are excluded, and the analysis is restricted to power-law FG expansions.

\section{Asymptotic analysis and renormalization\label{sec:asymptotics}}

To construct the extrinsic boundary counterterms of NMG, one first needs to solve the field equations~\eqref{eom2} order by order in the asymptotic expansion. The resulting relations for the FG coefficients are then used to evaluate the on-shell action and its variation near the conformal boundary. This will allow to determine the divergences in the different sectors of NMG. 
The analysis begins in Gauss-normal coordinates and then imposes the Fefferman-Graham gauge~\eqref{FGexp}, with the boundary located at $z=0$.

\subsection{Field equations}

 The NMG field equations~\eqref{eom2} can be decomposed using radial foliation. This is first performed in the Gauss-normal coordinates of~\eqref{GNcoord}. The independent components of the field equations can then be written as
\begin{subequations}
    \begin{align}
   \mathcal{E}_{z}^{z}	&=\sigma_{1}G_{z}^{z}+\lambda+\frac{\sigma_{2}}{\mu^{2}}\left(\frac{1}{2}(S_{i}^{i}S_{j}^{j}-S_{j}^{i}S_{i}^{j})+\nabla_{i}C_{z}^{\ iz}\right)\,, \\ \notag
\mathcal{E}_{j}^{i}	&=\sigma_{1}G_{j}^{i}+\lambda\delta_{j}^{i}+\frac{\sigma_{2}}{2\mu^{2}}\left(2\left[S_{k}^{i}S_{j}^{k}+S_{z}^{i}S_{j}^{z}-S_{j}^{i}\left(S_{k}^{k}+S_{z}^{z}\right)+\nabla_{z}C_{j}^{\ z i} +\nabla_{k}C_{j}^{\ k i}\right.\right]\\
&\quad\left.+\, \delta_{j}^{i}\left[S_{k}^{k}(S_{m}^{m}+2S_{z}^{z})-2S_{z}^{k}S_{k}^{z}-S_{m}^{k}S_{k}^{m}\right]\right)\,,\\
\mathcal{E}_{z}^{i}	&=\sigma_{1}G_{z}^{i}+\frac{\sigma_{2}}{\mu^{2}}\left(S_{j}^{i}S_{z}^{j}-S_{k}^{k}S_{z}^{i}+\nabla_{z}C_{z}^{\ z i} + \nabla_{k}C_{z}^{\ k i}\right)\,.
\end{align}
\end{subequations}
 The Gauss--Codazzi relations express the bulk Schouten tensor in terms of intrinsic and extrinsic quantities. The asymptotic behavior of its components near the conformal boundary is computed explicitly in Appendix~\ref{sec:GN-FG}. On the other hand, the falloff of the different components of the Einstein tensor can be obtained by rewriting it in terms of the Schouten as $G^\mu_\nu = S^\mu_\nu - \delta^\mu_\nu
S$. Then, inserting the asymptotic expansion into the field equations, we obtain that the different components are given by  
\begin{subequations}
    \begin{align} 
    \mathcal{E}^z_z &=\mathcal{E}_{(0)}-\frac{z}{2\ell^{3}}\varpi_{+}\tr g_{(1)}-\frac{z^{2}}{8\ell^{4}}\Big(\varpi_{-}\left[({\rm tr}g_{(1)})^{2}-{\rm tr}g_{(1)}^{2}\right] \nonumber\\
    &-2\varpi_{+}\left[{\rm tr}g_{(1)}^{2}+({\rm tr}g_{(1)})^{2}-4{\rm tr}g_{(2)}-2\ell^{2}\mathcal{R}_{(0)}\right] \Big)+\mathcal{O}(z^{3})\,, \label{eomzz} \\
    \mathcal{E}^i_j &= \mathcal{E}_{(0)}\delta_{j}^{i}+\frac{\varpi_{+}}{2\ell^{3}}\left[\left(g_{(1)j}^{i}-\delta_{j}^{i}\tr g_{(1)}\right)z-\frac{1}{2\ell}\left(g_{(1)k}^{i}g_{(1)j}^{k}-\tr g_{(1)}^{2}\delta_{j}^{i}\right)z^{2}\right]+\mathcal{O}(z^{3})\,\label{eomij}  \,, \\
    \mathcal{E}_{i}^{z}&=\frac{\varpi_{+}}{2\ell^{3}}\left(\mathcal{D}_{(0)j}g_{(1)i}^{j}-\mathcal{D}_{(0)i}\tr g_{(1)}\right)z^{2}+\mathcal{O}(z^{3})\,. \label{eomzi}
\end{align}
\end{subequations}
Here, the FG expansion~\eqref{FGexp} has been used by replacing $L$ with $\ell$, the effective AdS radius that appears in the field equations (see Sec.~\ref{sec:constant-curvature-spaces}). Additionally, indices are raised and lowered with the boundary metric $g_{(0)ij}$. 
Here, $D_{(0)i}$ denotes the covariant derivative compatible with $g_{(0)ij}$ and $\mathcal{R}_{(0)}$ is its associated Ricci scalar. The definitions of $\mathcal{E}_{(0)}$ and $\varpi_{+}$ given in Eqs.~\eqref{Lambdaeffconstraint} and~\eqref{degcondition}, respectively, have been used. It is also defined the quantity $\tr g_{(1)}^2:=g^i_{(1)j}g^j_{(1)i}$. Boundary indices are henceforth raised and lowered with respect to the boundary metric $g_{(0)ij}$

To leading order in the asymptotic expansion, Eqs.~\eqref{eomzz} and~\eqref{eomij} imply $\mathcal{E}_0=0$. Consequently, Eq.~\eqref{Lambdaeffconstraint} not only fixes the effective AdS radius of maximally symmetric spaces, but also defines the curvature radius for asymptotically AdS spacetimes in NMG. Additionally, away from the degenerate point $(\varpi_{+}\neq0)$, the next-to-leading-order term of Eq.~\eqref{eomzz} implies $\tr g_{(1)}=0$. Then, the same order in Eq.~\eqref{eomij} fixes $g_{(1)ij}=0$. To quadratic order in the holographic coordinate, Eq.~\eqref{eomzz} yields exactly the same condition found in Eq.~\eqref {trg2condition} but in this case with the effective cosmological constant determined by solving $\mathcal{E}_0=0$. Thus, as long as $\varpi_{\pm}\neq0$, the asymptotic solutions to the field equations yield 
\begin{align}\label{g-solNMG}
   \mathcal{E}_0 = 0\,, \qquad g_{(1)ij} = 0\,, \qquad \tr g_{(2)} = -\frac{\ell^2}{2}\,\mathcal{R}_{(0)}\,.
\end{align}
This equation contains Brown-Henneaux boundary conditions provided a suitable gauge fixing~\cite{Ciambelli:2020ftk,Ciambelli:2022vot}. Accordingly, the asymptotic solutions~\eqref{g-solNMG} are henceforth referred to as Brown--Henneaux boundary conditions  following Refs.~\cite{Ayon-Beato:2009cgh,Giribet:2010ed,Alishahiha:2010bw,Kwon:2011jz,Cunliff:2013en}.

On the other hand, at the degenerate point, i.e., $\varpi_{+}=0$, the equations of motion no longer require $g_{(1)ij}$ to vanish. This mode turns into new independent asymptotic data. The order $z^2$ in Eq.~\eqref{eomzz} imposes the algebraic constraint 
\begin{align}\label{trg1}
\tr g_{(1)}^2 = (\tr g_{(1)})^2\; \quad  \to \quad g^k_{(1)i}g_{(1)jk}=\tr g_{(1)}\,g_{(1)ij} \,,
\end{align}
where the Cayley-Hamilton relation for $g_{(1)ij}$ has been used in the last equality.

The solutions relevant to the asymptotic analysis below split into three sectors. The first one is the Einstein sector, characterized by constant-curvature spaces satisfying the properties of Sec.~\ref{sec:constant-curvature-spaces}. It contains AdS$_3$ and its locally equivalent quotients, including the BTZ black hole~\cite{Banados:1992wn,Banados:1992gq}. These geometries have vanishing conserved charges at the critical point, $\varpi_-=0$, for Brown-Henneaux boundary conditions. The second sector consists of non-Einstein spaces with standard AdS asymptotics. Although their Schouten tensor is not constant, the generic asymptotic equations impose $g_{(1)ij}=0$, while their metric has the same falloff as that of the Einstein sector. The third sector consists of non-Einstein spaces at the degenerate point with relaxed AdS boundary conditions. In this case, the Fefferman--Graham expansion may contain $g_{(1)ij} \neq 0$. This class of solutions contains the static and rotating hairy black holes of Refs.~\cite{Oliva:2009ip,Giribet:2009qz,Bergshoeff:2009aq}. Notice that such geometries may be conformally flat, and hence have a vanishing Cotton tensor and, at the same time, be non-Einstein. The presence of $g_{(1)ij}$ is crucial to the renormalization problem because it generates additional divergent contributions to the action.

\subsection{Asymptotic expansion of the action and variations thereof}

Having solved the field equations asymptotically, the intention now is to determine the divergences of the on-shell action. To this end, one expands the NMG Lagrangian defined in Eq.~\eqref{INMG} near the conformal boundary as 
\begin{align}\label{LagFG}
    \kappa^{-1}\Lag_{\rm NMG} = \Lag_{(0)}+z\Lag_{(1)}+z^{2}\Lag_{(2)}+\mathcal{O}(z^{3})\,.
\end{align}
Since different sectors of the NMG parameter space will be explored, the general asymptotic expansion of the action is presented. 

The asymptotic solutions appropriate to each sector are imposed after presenting the general expansion. Thus, the first terms in the FG expansion  of the Lagrangian are given by 
\begin{subequations} \label{Lag-coeff}
    \begin{align}
   \Lag_{(0)}	&= -2\mathcal{E}_0 -\frac{4\varpi_{-}}{\ell^2}\,, \qquad \Lag_{(1)}	=\frac{2\varpi_{-}\tr g_{(1)}}{\ell^{3}}\,, \label{L0L1} \\
    \Lag_{(2)}	&=\frac{\varpi_{+}}{4\ell^{4}}\left([\tr g_{(1)}]^{2}-\tr g_{(1)}^{2}\right)
	-\frac{\varpi_{-}}{2\ell^{4}}\left([\tr g_{(1)}]^{2}+2\tr g_{(1)}^{2}-2\left(2\tr g_{(2)}+\ell^{2}\mathcal{R}_{(0)}\right)\right) \label{L2} \,. 
\end{align}
\end{subequations}
Notice that, for Einstein spaces and non-Einstein spaces outside the degenerate and critical points, the only nonvanishing term is $\Lag_{(0)}$. This term contributes with a volume divergence that needs to be renormalized. More explicitly, the volume element behaves as
\begin{equation} \label{sqrtdetgexp}
    \sqrt{|g|} = \frac{\ell^3}{z^3} \sqrtgbdy \left[ 1 + \frac{z}{2 \ell} \tr g_{(1)} + \frac{z^2}{2 \ell^2} \left( \tr g_{(2)} + \frac{1}{4} (\tr g_{(1)})^2 - \frac{1}{2}\tr g_{(1)}^2   \right)   + \mathcal{O}(z^3) \right]\,,
\end{equation}
as $z\to0$. Thus, in order to obtain the asymptotic expansion of the action, one combines the coefficients~\eqref{Lag-coeff} with the volume element in Eq.~\eqref{sqrtdetgexp}, such that the bulk action takes the form 
\begin{equation}
    I_{\rm NMG} = \kappa \int_\mathcal{M} \diff{}^3x  \sqrtgbdy \left( \frac{\mathcal{L}_{(-3)}}{z^3} + \frac{\mathcal{L}_{(-2)}}{z^2} + \frac{\mathcal{L}_{(-1)}}{z} +\mathcal{O}(1)\right)\,. 
\end{equation}
where the coefficients $\mathcal{L}$ are defined by 
\begin{subequations}
    \begin{align}
        \mathcal{L}_{(-3)} &= \ell^3 \Lag_{(0)}\,, \qquad \mathcal{L}_{(-2)}= \frac{\ell^2}{2} (2\ell \Lag_{(1)} + \tr g_{(1)} \Lag_{(0)} )\,, \\
        \mathcal{L}_{(-1)}	&=\frac{\ell}{8}\left(\left[(\tr g_{(1)})^{2}-2\tr g_{(1)}^{2}+4\tr g_{(2)}\right]\Lag_{(0)}+4\ell\,\tr \,g_{(1)}\Lag_{(1)}+8\ell^{2}\Lag_{(2)}\right) \,,
    \end{align}
\end{subequations}
If $\varpi_{\pm}\neq0$,  the asymptotic solution to the NMG field equations implies that Eq.~\eqref{g-solNMG} must hold. Therefore, only $\mathcal{L}_{(-3)}$ is nonvanishing. This happens for both Einstein and non-Einstein spaces, away from the degenerate and critical points. On the other hand, if $\varpi_{+}=0$ while $\varpi_{-}\neq0$, in general, $g_{(1)ij}\neq0$ and extra divergencies might appear. 

In addition to rendering the on-shell action finite, the boundary terms must ensure a differentiable action compatible with the prescribed asymptotic boundary conditions. Canceling the divergences of the on-shell action does not, by itself, establish this property. One must also verify that its on-shell variation is finite and can be expressed in terms of variations of the prescribed holographic sources, without contributions involving variations of the asymptotic data that are unfixed. The resulting boundary variation then vanishes when the sources are held fixed, establishing the corresponding Dirichlet variational principle.
For the constant and nonconstant curvature sectors considered here with $g_{(1)ij}=0$, the boundary metric $g_{(0)ij}$ is the only independent holographic source, and the finite coefficient of $\delta g_{(0)ij}$ determines its associated stress tensor. At the degenerate point, however, the asymptotic field equations allow a nonvanishing $g_{(1)ij}$. One therefore considers a generalized Dirichlet problem in which both $g_{(0)ij}$ and $g_{(1)ij}$ are prescribed, subject to the asymptotic constraints, and requires a finite response conjugate to each source. The sectors are analyzed separately below.

\subsubsection{Constant-curvature spaces}

First, the renormalization of NMG for constant-curvature spaces with Brown-Henneaux boundary conditions. These spaces, in three dimensions, represent Einstein spaces. In that case, the Ricci scalar and Schouten tensors reduce to 
\begin{equation}\label{Einstein-spaces}
    R= - \frac{6}{\ell^2}\,, \qquad S^\mu_\nu = - \frac{1}{2\ell^2} \delta^\mu_\nu\,,
\end{equation}
while the Cotton tensor vanishes. Using Eq.~\eqref{Einstein-spaces} alongside the asymptotic solutions~\eqref{g-solNMG}, the bulk on-shell NMG action turns out to be
\begin{align}\label{INMGE}
    I_{\rm NMG}\big|_{\rm E} = -\frac{4\kappa\varpi_{-}}{\ell^2}\int_{\mathcal{M}}\diff{^3x}\, \sqrtgbdy \left[ \frac{\ell^3}{z^3}   + \frac{\ell}{2z} \tr g_{(2)}   + \mathcal{O}(z) \right]\,.
\end{align}
The overall coefficient is proportional to $\varpi_{-}$. Therefore, the bulk on-shell action vanishes at the critical point for constant-curvature spacetimes with Brown-Henneaux boundary conditions. Away from criticality, the first term of  Eq.~\eqref{INMGE} produces the usual volume divergence, whereas the second term does not contribute to any divergence whatsoever since it is proportional to the Euler density. 

As in Einstein-AdS gravity, the NMG action and variations thereof for constant-curvature spaces  can also be renormalized by introducing a purely extrinsic boundary term. At the same time, it should appear endowed with a well-posed variational principle for Dirichlet boundary conditions for $g_{(0)}$ and, therefore, compatible with a holographic picture. As a matter of fact, this can be realized by considering $I_{\rm ren}=I_{\rm NMG} +I_K$, where the former is defined in Eq.~\eqref{INMG}, while the latter is
\begin{align}\label{KT-NMG-E}
     I_{K}=-\kappa\varpi_{-} \int_{\partial\mathcal{M}} \diff{}^2x \sqrt{|h|} \; K \,.    
\end{align}
Indeed, for $\mu\to\infty$, the coefficient of $I_K$ approaches $\kappa \sigma_1$, reproducing the half-GHY prescription of Sec.~\ref{sec:3D-Einstein}. Performing arbitrary variations of $I_{\rm ren}$ produces
\begin{equation} \label{deltaIrennshell}
    \delta I_{\rm ren}|_{\rm E}  = \int_{\partial \mathcal{M}} \diff{}^2x \left[ \sqrt{|h|}\; n_\mu \Theta^\mu - \kappa \varpi_{-} \delta \left( \sqrt{|h|} K \right) \right]\,, 
\end{equation}
with $\Theta^\mu$ being defined in Eq.~\eqref{Thetamu2} and $n_\mu = -N \delta_\mu^z$ is the unit normal to the constant-$z$ hypersurfaces, with $n^\mu = - N^{-1} \delta^\mu_z$. Particularizing the contributions to Einstein spaces, one obtains
\begin{equation}\label{nmuThetamuEinstein}
    n_\mu \Theta^\mu |_{\rm E} = \kappa \varpi_{-} n_\mu \delta \Gamma^\sigma_{\nu \rho} g^{\rho \alpha} \delta^{\mu \nu}_{\sigma \alpha} = \kappa\varpi_{-}\left(2\delta^{i}_{j}\delta K^{j}_{i}+K^i_j(h^{-1}\delta h)^j_i\right) \,.
\end{equation}
This variation resembles that of Einstein gravity but with a modified overall coefficient, which is nonvanishing as long as $\varpi_{-}\neq0$. Thus, once the Einstein condition~\eqref{Einstein-spaces} and the asymptotic solutions in Eq.~\eqref{g-solNMG} are taken into account, the variation of the full action at the conformal boundary is exactly that of Eq.~\eqref{deltaIEtilde}, but with a different overall coefficient. The corresponding holographic stress tensor can be read off from that relation by replacing $\tau_{\rm (E)}^{ij}$ by $\tau^{ij}$ in Eq.~\eqref{FG-GHY}, where
\begin{equation}\label{tauijNMG}
    \tau^{ij} = -\frac{2\kappa \varpi_-}{\ell }  \left(g_{(2)}^{ij} -g_{(0)}^{ij} \tr g_{(2)} \right) \,.
\end{equation}
%\cc{while} the total variation \cc{\sout{on the right-hand side of Eq.~\eqref{tauijNMG}}} has been neglected in the definition of $\tau^{ij}_{(E)}$, since it represents the variation of the two-dimensional Euler density\cc{\sout{ and therefore it does not contribute to the on-shell variation}}. 
Equation~\eqref{tauijNMG} is finite and expresses the on-shell variation entirely in terms of the boundary source $g_{(0)ij}$. Consequently, the variation vanishes when Dirichlet boundary conditions, $\delta g_{(0)ij}=0$, are imposed. The boundary term~\eqref{KT-NMG-E} therefore defines a finite and well-posed Dirichlet variational principle for constant-curvature solutions with Brown--Heanneaux boundary asymptotics.

\subsubsection{Nonconstant-curvature spaces with \texorpdfstring{$g_{(1)ij}=0$}{g1ijequal0}}

For nonconstant-curvature solutions, that is, the branch defined by Eq.~\eqref{g-solNMG}, the Schouten tensor is not proportional to the identity and the Cotton tensor does not vanish. In that case, the on-shell variation of $I_{\rm NMG}+I_K$, in Gauss-normal coordinates, is given by  \begin{equation} \label{bdytermnonconstant}
    n_\mu \Theta^\mu = 2 N \delta \Gamma^k_{ij} h^{j\ell} E^{zi}_{k\ell} + 4 \delta K^i_j E^{zj}_{zi} + \left( 2 K^j_k E^{zk}_{zi} - \frac{\kappa \sigma_2}{\mu^2} N C_i{}^{zj} \right) (h^{-1}\delta h)^i_j \,.
\end{equation}
Notice that the relevant components of the $E$-tensor are 
\begin{equation}
    E^{zi}_{zj} = \kappa \left[\frac{\sigma_1}{2} \delta^i_j - \frac{\sigma_2}{2\mu^2} \left( S^i_j + \delta^i_j S^z_z - \delta^i_j S \right) \right]\,, \qquad E^{z \ell}_{kj} =  \frac{\kappa \sigma_2}{ \mu^2 }  \delta^\ell_{[k} S^z_{j]} \,. 
\end{equation}

The proper use of Gauss-Coddazi relations in Appendix~\ref{GNcoord}, then gives
\begin{align}
    \delta I_{\rm ren} &= \kappa \int_{\partial \mathcal{M}} \diff{}^2 x\, \sqrt{|h|} \left\{ \sigma_1 \left[ \delta^i_j \delta K^j_i + \bar{K}^i_j (h^{-1}\delta h)^j_i \right] \right. \nonumber +\frac{\sigma_2}{\mu^2} \left[ \left( \left[ \frac{1}{2\ell^2}+2 S^k_k\right] \delta^i_j - 2 S^i_j \right) \delta K^j_i\right. \\
    &\left. \left. +\left( \frac{1}{N} \partial_z S^i_j - 2K^i_k S^k_j + SK^i_j + 2\mathcal{D}^i \mathcal{U}_j + \left[ \frac{K}{2\ell^2} - \mathcal{D}_k \mathcal{U}^k \right]\delta^i_j\right) (h^{-1} \delta h)^j_i \right] \right\}
\end{align}
where the quantities $\mathcal{U}_i$, and $\bar{K}^i_j$ are defined as
\begin{align}
    S=S^z_z+S^i_i\,, \quad \mathcal{U}_{i}=\mathcal{D}_{j}K_{i}^{j}-\mathcal{D}_{i}K\,, \quad \bar{K}_{j}^{i}=K_{j}^{i}-\frac{1}{2}\delta_{j}^{i}K \,.
\end{align}
Plugging in the FG expansion into this expression, one can verify that, indeed, all divergent contributions are canceled. What remains is holographic, as it is the finite part and expressible in terms of the holographic source
\begin{equation}
    \delta I_{\rm ren} = \frac12 \int_{\partial \mathcal{M}} \diff{}^2x \, \sqrtgbdy\, \tau^{ij} \delta g_{(0)ij} \,,
\end{equation}
where $\tau^{ij}$ is the stress tensor in Eq.~\eqref{tauijNMG}.

\subsubsection{Degenerate case \texorpdfstring{$g_{(1)ij}\neq0$}{}}

The field equations allows for a nontrivial $g_{(1)ij}$ mode when $\varpi_{+}=0$. This term in the metric, linear in $z$, modifies the expansion of the action. Using the fact $\varpi_-	=2\sigma_{1}$ at the degenerate point, together with $\mathcal{E}_0 = 0$, the first coefficients in the expansion of the NMG Lagrangian in Eq.~\eqref{LagFG} are
\begin{subequations}
    \begin{align}
    \Lag_{(0)}	&=-\frac{8\sigma_{1}}{\ell^{2}}\,, \qquad \Lag_{(1)}=\frac{4\sigma_{1}}{\ell^{3}}\tr g_{(1)}\,, \\
    \Lag_{(2)}	 &=\frac{2\sigma_{1}}{\ell^{4}}\left(2\tr g_{(2)}+\ell^{2}\mathcal{R}_{(0)}-\frac{3}{2}\tr g_{(1)}^2\right)\,. 
\end{align}
\end{subequations}
Then, combining these coefficients with the asymptotic form of the volume element~\eqref{sqrtdetgexp}, one finds
\begin{equation}
    I_{\rm NMG}|_{\rm deg} = \kappa \sigma_1 \int_{\mathcal{M}} \diff{}^3 x \sqrtgbdy \left[-\frac{8\ell}{z^3} + \frac{2\ell}{z} \mathcal{R}_{(0)} +\mathcal{O}(z) \right] \,. 
\end{equation}
Despite the weakened falloff in the FG expansion, the bulk action does not contain new divergences associated with it. The $z^{-1}$ term produces the same logarithmic term found in the $g_{(1)ij}=0$ case, which does not contribute to the dynamics because of its topological nature. 

Contrary to the case with $g_{(1)ij} =0$, the determinant of the metric~\eqref{sqrtdethexp} contains a linear term in $z$. In turn, the components of the extrinsic curvature behave as~\eqref{K-FG}. The boundary term in Eq.~\eqref{KT-NMG-E} therefore has the asymptotic form 
\begin{equation} \label{Ikdegenereque}
    I_K = -\kappa\varpi_{-} \int_{\partial\mathcal{M}} \diff{}^2 x \sqrtgbdy \left[ \frac{2\ell}{z^2} + \frac{\tr g_{(1)}}{2z} + \mathcal{O}(z)\right]\,.
\end{equation}
The first term cancels the leading volume divergence of the bulk action after radial integration. The second produces a residual divergence proportional to $\tr g_{(1)}$. To cancel this contribution, the following counterterm is proposed 
\begin{equation} \label{GBdyaction}
    I_{\rm ct} =  \kappa \varpi_-\int_{\partial \mathcal{M}} \diff{}^2x \,\sqrt{|h|} \left(\frac{1}{\ell}  - \frac{\ell}{2} \delta^{ik}_{jl} K^j_i K^l_k  \right)\,.
\end{equation}
The only divergent term in the boundary density of $I_{\rm ct}$ is $\kappa \varpi_- \sqrtgbdy\, \tr g_{(1)}/(2z)$, which cancels the corresponding contribution from $I_K$, while remaining are finite as $z\to0$.

Substituting the FG expansion into the on-shell variation of $I_{\rm NMG}+I_K+I_{\rm ct}$ and imposing the asymptotic field equations yields 
\begin{align}
    \delta &(I_{\rm NMG}+I_{\rm K} + I_{\rm ct})\big|_{\rm EOM} = \frac{\kappa \varpi_-}{4\ell} \int_{\partial \mathcal{M}} d^2 x\, \sqrtgbdy \left\{ \left[ \tr g_{(1)} g_{(1)}^{ij} - 4 g_{(2)}^{ij} \right.] \right. \nonumber \\
    &\left.  + \left( 2 \tr g_{(2)} - \ell^2 \mathcal{R}_{(0)} - \frac12 \tr g_{(1)}^2  \right) g_{(0)}^{ij} \right] \delta g_{(0)ij}  \left.-\left( g_{(1)}^{ij} - \tr g_{(1)} g_{(0)}^{ij}  \right) \delta g_{(1)ij} \right\} \,.
\end{align}
Equivalently, this defines the holographic responses through 
\begin{equation}
    \delta (I_{\rm NMG}+I_{\rm K} + I_{\rm ct})\big|_{\rm EOM} = \frac12 \int_{\partial \mathcal{M}} \diff{}^2 x\, \sqrtgbdy \left( T^{ij} \delta g_{(0)ij} + P^{ij}\delta g_{(1)ij} \right)\, .
\end{equation}
The responses conjugate to $g_{(0)ij}$ and $g_{(1)ij}$ are, 
\begin{subequations}
    \begin{align}
        T^{ij}	&=\frac{\kappa\varpi_-}{2\ell} \Big[4g_{(2)}^{ij} -\tr g_{(1)}g_{(1)}^{ij}+\Big(2 \tr g_{(2)}-\ell^{2}\mathcal{R}_{(0)}-\frac{1}{2} \tr g_{(1)}^{2}\Big)g_{(0)}^{ij}\Big]\,, \\
        P^{ij}	&=\frac{\kappa\varpi_-}{2\ell}\left[g_{(1)}^{ij}-\tr g_{(1)}g_{(0)}^{ij}\right]\,,
    \end{align} 
\end{subequations}
respectively. 

The on-shell variation identifies two boundary sources, $g_{(0)ij}$ and $g_{(1)ij}$, with conjugate responses $T^{ij}$ and $P^{ij}$, respectively. A well posed-variational principle for the relaxed branch therefore follows from Dirichlet boundary conditions on both sources, namely $\delta g_{(0)ij}=\delta g_{(1)ij}=0$. Restricting to $g_{(1)ij}=0$, and variations preserving this condition reduces the on-shell variation to the response conjugate to $g_{(0)ij}$. This structure is analogous to that encountered in Conformal Gravity~\cite{Grumiller:2013mxa}.

Holographic Ward identities and conserved charges can be obtained by assuming invariance under boundary diffeomorphisms of the form $\delta_\xi g_{(0)ij} =\Lie_\xi g_{(0)ij}$ and $\delta_\xi g_{(1)ij}=\Lie_\xi g_{(1)ij}$, where $\Lie_\xi$ is the Lie derivative along the generator $\xi=\xi^i\partial_i$. These requirements yield the holographic Ward identity
\begin{align}
    \mathcal{D}^{(0)}_j\left( T^{ij} + P^{jk}g^i_{(1)k} \right) = \frac{1}{2}\,P^{jk}\mathcal{D}^i_{(0)}g_{(1)jk}\,.
\end{align}
Additionally, the conserved charges associated with the asymptotic Killing vector are
\begin{align}\label{chargedegTijPij}
    Q[\xi] = \int_\Sigma \diff{x}\sqrt{|\gamma|} \, u^i\left(T_{ij} + P^{k}{}_i \, g_{(1)jk}\right)\,\xi^j \,.
\end{align}
Thus, in the degenerate case, the weakened AdS boundary conditions nontrivially modify the conserved charges, similar to what happens in four-dimensional Conformal Gravity~\cite{Grumiller:2013mxa}. 

\subsection{Comparison to Hohm--Tonni stress tensor} 

The auxiliary-field prescription of Ref.~\cite{Hohm:2010jc} provides an alternative construction of the holographic stress tensor. It introduces an independent symmetric tensor $f_{\mu \nu}$ that couples linearly to the Einstein tensor and enters quadratically through an algebraic term. In the conventions adopted here, the corresponding bulk action is 
\begin{equation}
    I_{\rm aux} = \kappa \int_{\mathcal M} \diff{}^3x \sqrt{|g|} \left[ \sigma_1 R - 2\lambda + f^{\mu \nu} G_{\mu \nu} + \frac{\sigma_2\mu^2}{4} \left( f_{\mu \nu} f^{\mu \nu} - f^2 \right) \right]\,,
\end{equation}
where $f = g^{\mu \nu}f_{\mu \nu}$. Variation with respect to $f^{\mu \nu}$ gives the algebraic equation for the auxiliary field $G_{\mu \nu} + \tfrac12 \sigma_2 \mu^2 (f_{\mu \nu}-fg_{\mu \nu})=0$, whose solution is 
\begin{equation}\label{auxiliarySchouten}
    f_{\mu\nu} =-\frac{2\sigma_2}{\mu^2}S_{\mu\nu}\,.
\end{equation}
Plugging this expression into $I_{\rm aux}$ reproduces the curvature squared term in the NMG action. Taking the metric and auxiliary tensor as independent variables gives a coupled system of equations containing at most second derivatives. This formulation provides a direct construction of the generalized Gibbons--Hawking term by treating the boundary values of the metric and auxiliary field as Dirichlet data~\cite{Hohm:2010jc}.

Introducing the tangential projector $h^\mu_\nu=\delta^\mu_\nu-n^\mu n_\nu$, where $n_\mu$ is the unit normal to the constant-$z$ hypersurfaces defined above, while the tangential, mixed, and normal projections of the auxiliary tensor are defined by 
\begin{equation}\label{auxiliaryProjections}
    \hat f^{ij}=f^{\mu\nu}h_\mu^i h_\nu^j\,,
    \qquad
    \hat h^i=f^{\mu\nu}h_\mu^i n_\nu\,,
    \qquad
    \hat s=f^{\mu\nu}n_\mu n_\nu\,,
\end{equation}
respectively, with $\hat f= h_{ij} \hat f^{ij}$ denoting the tangential trace. Tangential indices are raised and lowered with $h_{ij}$, and $\mathcal D_i$ denotes its compatible covariant derivative. 

The action supplemented by the generalized Gibbons--Hawking term and the intrinsic counterterm is
\begin{align}\label{ITonniHohm}     I_{\rm HT}     ={}&I_{\rm         NMG}  -   \kappa\int_{\partial\mathcal M}\diff{}^2x\,\sqrt{|h|}     \left(         2\sigma_1K+\hat f^{ij}K_{ij}-\hat f K     \right) -\frac{2\kappa\varpi_-}{\ell}     \int_{\partial\mathcal M}\diff{}^2x\,\sqrt{|h|}\,. \end{align}
The term proportional to $\sigma_1$ is the Gibbons--Hawking contribution from the Einstein--Hilbert sector. The terms involving $\hat f^{ij}$ and $\hat f$ extend this boundary action to the curvature squared sector, while the final integral supplies the intrinsic counterterm.

The metric variation, with the mixed auxiliary field $f^\mu{}_\nu$ held fixed at the boundary, defines the renormalized stress tensor according to
\begin{equation}\label{variationHT}
    \delta I_{\rm HT}
    =
    \frac12\int_{\partial\mathcal M}\diff{}^2x\,\sqrt{|h|}
    \,T_{\rm ren}^{ij}\delta h_{ij}\,.
\end{equation}
The contribution from the bulk action and the generalized Gibbons--Hawking term is
\begin{align}\label{stressHT}
    T_{\rm HT}^{ij}
    =2\kappa\Bigg[
    &\sigma_1\left(K^{ij}-Kh^{ij}\right)
    -\frac12\hat f K^{ij}
    +\mathcal D^{(i}\hat h^{j)}
    -\frac{1}{2N}\partial_z\hat f^{ij}
    -K_k^{(i}\hat f^{j)k}
    +\frac12\hat s K^{ij}
    \nonumber\\
    &+h^{ij}\left(
        -\mathcal D_k\hat h^k
        -\frac12\hat s K
        +\frac12\hat f K
        +\frac{1}{2N}\partial_z\hat f
    \right)
    \Bigg]\,.
\end{align}
Including the intrinsic counterterm gives
\begin{equation}\label{Tijren}
    T_{\rm ren}^{ij}
    =
    T_{\rm HT}^{ij}
    -\frac{2\kappa\varpi_-}{\ell}h^{ij}\,.
\end{equation}
Substituting Eq.~\eqref{auxiliarySchouten} and the FG expansions into Eq.~\eqref{Tijren}, together with the asymptotic relations~\eqref{g-solNMG} for the nondegenerate branch, gives the on-shell variation in terms of the boundary source $g_{(0)ij}$. Taking the limit $z\to0$ yields
\begin{equation}\label{variationHTFG}
    \delta I_{\rm HT}
    =
    \frac{\kappa\varpi_-}{\ell}
    \int_{\partial\mathcal M}\diff{}^2x\,\sqrtgbdy
    \left(
        g_{(2)}^{ij}
        -g_{(0)}^{ij}\tr g_{(2)}
    \right)\delta g_{(0)ij}\,,
\end{equation}
variation that is holographic and defines the stress tensor at the conformal boundary. The method recovers the tensor, with the corresponding coupling, as obtained from the extrinsic prescription in Eq.~\eqref{tauijNMG}. This analysis proves explicitly the equivalence between our method and that of Ref.~\cite{Hohm:2010jc} for all spacetimes satisfying the asymptotic conditions~\eqref{g-solNMG}, including Einstein spaces and nonconstant-curvature spaces with $g_{(1)ij}=0$.

\section{Noether-Wald charges\label{sec:NW}}

Diffeomorphism invariance of the gravitational action~\eqref{INMG} under the flow generated by a Killing vector field $\xi=\xi^\mu\partial_\mu$ implies the conservation law $\nabla_\mu J^\mu = 0$ on-shell, where $J^\mu:=\Theta^\mu - \xi^\mu\Lag_{\rm NMG}$ is the Noether current. The Poincaré lemma states that the current should be locally a total derivative $J^\mu=\nabla_\nu q^{\mu\nu}$, whose prepotential remains ambiguous. However, in the Noether procedure, this tensor can be derived  as 
\begin{align}\label{Noetherprep}
    q^{\mu \nu} = -2 (E^{\mu \nu}_{\lambda \rho} \nabla^\lambda \xi^\rho + 2 \xi^\lambda \nabla^\rho E^{\mu \nu}_{\lambda \rho}) = -q^{\nu\mu}\,.
\end{align}
Then, conserved charges associated with the Killing vector $\xi$ can be computed by integrating the Noether prepotential over a codimension-2 hypersurface. In the case of black holes, the Bekenstein-Hawking entropy is obtained by integrating the Noether charge at the bifurcating Killing horizon~\cite{Wald:1993nt,Iyer:1994ys,Wald:1999wa}. In the case of asymptotic charges, they can be obtained from the boundary integral
\begin{align}\label{QNW}
    \mathcal{Q}[\xi]= \int_{\Sigma_\infty}\left(q^{\mu\nu} - 2\xi^{[\mu}n^{\nu]}\,\mathcal{B}\right)\diff{\Sigma_{\mu\nu}}\,,
\end{align}
where $n=n^\mu\partial_\mu$ is a spacelike unit normal to the codimension-1 hypersurface that defines radial foliation and $\diff{\Sigma_{\mu\nu}}$ is the oriented volume element of the codimension-2 asymptotic boundary $\Sigma_\infty$. Additionally, $\mathcal{B}$ is the boundary term required to define finite conserved charges, as well as to establish a well-posed variational principle; for $\varpi_\pm\neq0$, the boundary contribution is given by Eq.~\eqref{KT-NMG-E}, whereas Eq.~\eqref{GBdyaction} applies at the degenerate point $\varpi_+=0$, Inserting the $E$-tensor~\eqref{Etensor} into the Noether prepotential~\eqref{Noetherprep} yields
\begin{align} \label{eq:pnoether}
    q^{\mu\nu} =-2\kappa\left[ \frac{1}{2}\left(\sigma_1 +\frac{\sigma_2}{\mu^2}S\right)\delta^{\mu\nu}_{\lambda\rho}-\frac{2\sigma_2}{\mu^2}S^{[\mu}_{[\lambda}\delta^{\nu]}_{\rho]} \right]\nabla^\lambda\xi^\rho +\frac{2\kappa \sigma_2}{\mu^2}\xi^\lambda C_{\lambda}{}^{\mu\nu}\,. 
\end{align}

In the case of constant-curvature spaces, the Schouten tensor is $S^\mu_\nu=-\tfrac{1}{2\ell^2}\delta^\mu_\nu$, the Cotton tensor vanishes, and the Noether prepotential~\eqref{eq:pnoether} becomes
\begin{align}
    q^{\mu\nu}\big|_{\rm E} = -2\kappa\varpi_-\nabla^{[\mu}\xi^{\nu]}\,.
\end{align}
Notice that, at the critical point, the Noether-Wald charges vanish for all Einstein spaces with Brown-Henneaux boundary conditions. This is similar to what happens with Critical Gravity in four and six dimensions~\cite{Lu:2011zk,Anastasiou:2017rjf,Anastasiou:2021tlv}. For nonconstant-curvature asymptotically AdS spaces, on the other hand, the Schouten tensor is not necessarily constant. Then, the relevant components of the Noether prepotential in Gauss-normal coordinates Eq.~\eqref{GNcoord} are given by
\begin{align}
     %q^{zi} &= -\kappa\sigma_1\nabla^z\xi^i + \frac{\kappa\sigma_2}{\mu^2}\left(S^i_j - \delta^i_j S^k_k \right)\nabla^z\xi^j - \frac{2\kappa\sigma_2}{\mu^2}\left(S^z_j\nabla^{[i}\xi^{j]} - \xi^j C_{j}{}^{zi} \right)
     q^{zi} &= -\kappa\left[\sigma_1 + \frac{\sigma_2}{\mu^2}\left(S-S^z_z \right) \right]\nabla^z\xi^i + \frac{\kappa\sigma_2}{\mu^2}S^i_j\nabla^z\xi^j - \frac{2\kappa\sigma_2}{\mu^2}\left(S^z_j\nabla^{[i}\xi^{j]} - \xi^j C_{j}{}^{zi} \right)\,, \label{qzi1}
\end{align}
where the Killing vector field is taken to be tangent to the constant radial coordinate hypersurfaces, $\xi=\xi^i\partial_i$. Then, the falloff of the covariant derivative of the Killing vector towards the asymptotic boundary off-shell is
\begin{subequations}
\begin{align}
    \nabla^{[z}\xi^{i]} &= -\frac{z}{\ell^2}\xi^i + \frac{z^2}{2\ell^3}\left( g_{(1)j}^i \xi^j+\ell\partial_z \xi^i \right)+ \frac{z^3}{2\ell^4} \left(2 g_{(2)j}^i  - g_{(1)k}^i g_{(1)j}^k \right) \xi^j+ \mathcal{O}(z^4)\,. 
\end{align}    
\end{subequations}
Expanding the Noether prepotential asymptotically towards the conformal boundary when $\varpi_\pm\neq0$ and imposing the Brown-Henneaux boundary conditions~\eqref{g-solNMG}] yields
\begin{align}
    \mathcal{Q}[\xi] = \frac{2\kappa\varpi_-}{\ell}\int_{\Sigma_\infty}\diff{x}\sqrt{|\gamma|}\,u_{i}\xi^{j} \left( g^i_{(2)j} + \frac{\ell^2}{2} \delta^i_j  \mathcal{R}_{(0)} \right)\,,
\end{align}
where $u_{i}$ denotes its unit timelike normal. This result matches exactly the conserved charges obtained via the holographic stress tensor in Eq.~\eqref{chargetauijNMG}. 

On the other hand, in the degenerate case ($\varpi_+=0$), and after the contribution of the extrinsic counterterm in Eq.~\eqref{GBdyaction} is taken into account, the conserved charge associated with an asymptotic Killing vector $\xi^{i}$ 
\begin{align} \label{Chargefromaction}
    \mathcal{Q}[\xi] &=\frac{2\kappa\varpi_-}{\ell}\int_{\Sigma_{\infty}}dx\,\sqrt{|\gamma|}\,u_{i}\xi^{j}\Bigg[g_{(2)j}^{i}-\frac{1}{2}{\rm tr}g_{(2)}\delta_{j}^{i} \nonumber \\
    &-\frac{1}{4}{\rm tr}g_{(1)}\left(g_{(1)j}^{i}-\frac{1}{2}{\rm tr}g_{(1)}\delta_{j}^{i}\right)+\frac{\ell^{2}}{4}\delta_{j}^{i}\mathcal{R}_{(0)}\Bigg] \,.
\end{align}
Using Eq.~\eqref{trg1}, one can show that the resulting charge agrees with Eq.~\eqref{chargedegTijPij} at the degenerate point. 

\section{Black holes and AdS waves: Conserved charges} \label{Sec.BHandConservedCharges}

Although the NMG field equations allow constant-curvature black holes, there are interesting asymptotically AdS black hole and AdS wave solutions which depart from this condition. The evaluation of conserved charges for different sectors of NMG provides a concrete application of the proposed boundary terms and permits a direct comparison with existing methods in the literature. 

\subsection{The BTZ black hole}

The BTZ black hole~\cite{Banados:1992wn,Banados:1992gq} is locally AdS with constant curvature given by $R^{\mu\nu}_{\lambda\rho}=-\tfrac{1}{\ell^2}\delta^{\mu\nu}_{\lambda\rho}$. This solution is described by the line element
\begin{equation}
    \diff{s^2} = -f(r)\,\diff{}t^{2} + \frac{\diff{}r^{2}}{f(r)} + r^{2}\bigl(N(r)\,\diff{t} + \diff{\varphi}\bigr)^{2},
\end{equation}
where the metric functions that solve the field equations are
\begin{equation}
    \label{eq:placeholder_label}
    f(r) = -8mG + \frac{16 j^{2} G^{2}}{r^{2}} + \frac{r^{2}}{\ell^{2}}
    \qquad \text{and} \qquad
    N(r) = - \frac{4 j G}{r^{2}} \,,
\end{equation}
with $m$ and $j$ being integration constants. The singularity at $r=0$ is hidden behind an inner and outer event horizons at $r=r_\pm$, characterized by the real positive roots of the polynomial $f(r_\pm)=0$, i.e.
\begin{equation}
    r_{\pm} = 2\ell \sqrt{G} \left[ m \pm \sqrt{m^{2} - j^{2}/\ell^{2}} \right]^{\frac{1}{2}}\,.
\end{equation}

The conserved charges associated with the Killing vectors $\partial_t$ and $\partial_\varphi$ can be computed using Eq.~\eqref{QNW}, giving 
\begin{subequations}\label{NWBTZ}
\begin{align}
    \mathcal{Q}[\partial_t] &:= M_{\rm BTZ} = m\left(\sigma_1 - \frac{\sigma_2}{2\mu^2\ell^2} \right) = m\varpi_-\,,\\
    \mathcal{Q}[\partial_\varphi] &:= \, J_{\rm BTZ}  \,= j\left(\sigma_1 - \frac{\sigma_2}{2\mu^2\ell^2} \right) = j\varpi_-\,.
\end{align}
\end{subequations}
This result matches that in Ref.~\cite{Hohm:2010jc} once the value $(\sigma_1,\sigma_2)=(1,-1)$ is fixed according to their conventions and, on the other hand, it also matches the result of Ref.~\cite{Clement:2009gq} using different methods for $(\sigma_1,\sigma_2)=(1,1)$. The entropy can be obtained via the Wald formalism, giving $S=\tfrac{A\varpi_-}{4G}$~\cite{Clement:2009gq}. The conserved charges in Eq.~\eqref{NWBTZ} satisfy the first law of thermodynamics for this higher-curvature corrected entropy.

\subsection{Nonconstant-curvature AdS waves}

AdS waves represent an exact class of analytical solutions of General Relativity with negative cosmological constant in $D\geq4$, first found in Refs.~\cite{Garcia:1981ads,Salazar:1983ads,Ozsvath:1985qn}. In three dimensions, however, AdS waves are diffeomorphic to global AdS in Einstein gravity~\cite{Ayon-Beato:2005gdo}. This is no longer true in the presence conformally coupled scalar fields, in Topologically Massive Gravity, or in NMG, where nonconstant curvature AdS waves have been found analytically~\cite{Ayon-Beato:2009cgh,Ayon-Beato:2005gdo}. They can be described by the metric in the Kerr-Schild form
\begin{subequations}\label{AdSwave}
\begin{align}
    \diff{s^2} = \left(g_{\mu\nu}^{\rm AdS} - F\,k_\mu k_\nu\right)\diff{x^\mu}\diff{x^\nu} = \frac{\ell^2}{z^2}\left( \diff{z^2} - 2\diff{u}\diff{v}  -F(u,z)\diff{u^2} \right) \,,
\end{align}
where $k=(\ell/z)\partial_u$ is a null geodesic vector. Inserting this ansatz into the NMG equations~\eqref{eom2}, one finds that the solution is given by~\cite{Ayon-Beato:2009cgh}
\begin{align}
    F(u,z) = F_+(u)z^{1+\ell\mu\sqrt{\sigma_2\varpi_+}} + F_-(u)z^{1-\ell\mu\sqrt{\sigma_2\varpi_+}}\,,
\end{align}
\end{subequations}
where $F_\pm(u)$ are arbitrary functions of the null coordinate $u$, while the homogeneous and quadratic order in $z$ functions were eliminated using the coordinate transformation discussed in extenso in Ref.~\cite{Ayon-Beato:2005gdo}. The solution has an asymptotically AdS behavior with Brown-Henneaux boundary conditions if $F_-(u)=0$, $\sigma_2=1$, and $\varpi_+>0$~\cite{Ayon-Beato:2009cgh}.

The solution in Eq.~\eqref{AdSwave} has a null Killing vector field given by $\xi=\partial_v$. Indeed, the conserved charges~\eqref{QNW} associated with the latter are finite but equal to zero. This resembles the behavior of ADT~\cite{Abbott:1981ff,Deser:2002rt,Deser:2002jk} and topologically renormalized Noether-Wald charges~\cite{Giribet:2018hck} of its four-dimensional counterpart in quadratic gravity~\cite{Giribet:2018hck}. 

\subsection{Hairy black hole}

At the degenerate point, i.e., $\varpi_+=0$, NMG admits a nonconstant-curvature black hole with relaxed AdS boundary conditions studied in Refs.~\cite{Oliva:2009ip,Giribet:2009qz}. The solution is given by
\begin{align}
    \diff{s^2} = - N(r)f(r)\diff{t^2} + \frac{\diff{r^2}}{f(r)} + r^2\left(\diff{\varphi} + N^\varphi(r)\diff{t} \right)^2\,,
\end{align}
where the metric functions are given by
\begin{align}\notag
    N(r)&=\left[1+\frac{b l^{2}}{4H(r)}(1-\eta)\right]^{2}\,, \quad
    N^{\varphi}(r)=-\frac{a}{2r^{2}}\left[4Gm-bH(r)\right]\,,\\
    f(r)&=\frac{H^{2}(r)}{r^{2}}\left[\frac{H^{2}(r)}{l^{2}}+\frac{b}{2}(1+\eta)H(r)+\frac{b^{2}l^{2}}{16}(1-\eta)^{2}-4mG\eta\right], \\
    H(r) &:= \left[ r^{2} - 2mGl^{2}(1 - \eta) - \frac{b^{2}l^{4}}{16}(1 - \eta)^{2} \right]^{\frac{1}{2}} \,, \quad \eta = \sqrt{1 - \frac{a^2}{\ell^2}}\,, \notag
\end{align}
with $m$, $a$, and $b$ being integration constants. Indeed, the $b$-mode turns on $g_{(1)ij}$, which produces a weakened AdS falloff. The Noether-Wald charges associated with the Killing vector fields $\partial_t$ and $\partial_\varphi$ give
\begin{subequations}\label{chargesOTT}
\begin{align}
    \mathcal{Q}[\partial_t] &:= M_{\rm OTT} = m + \frac{b^2\ell^2}{16G}\,, \\ \mathcal{Q}[\partial_\varphi] &:= J_{\rm OTT} = \left( m + \frac{b^2\ell^2}{16G} \right)a\,,
\end{align}    
\end{subequations}
respectively, in full agreement with the result obtained via the Abbott-Deser-Tekin~\cite{Abbott:1981ff,Deser:2002jk} method in Refs.~\cite{Oliva:2009ip,Giribet:2009qz} and the Tonni-Hohm quasilocal stress tensor with suitable counterterms for the weakened AdS falloff as done in Ref.~\cite{Giribet:2010ed}. Hence, the extrinsic boundary terms proposed here correctly reproduce the conserved charges previously found for the degenerate and non-degenerate cases.

\section{Conclusions\label{sec:conclusions}}

An extrinsic renormalization prescription for asymptotically AdS solutions of NMG has been developed. The construction is motivated by the fact that, while the renormalization of the Brown-Henneaux sector of NMG is well understood using intrinsic counterterms and the auxiliary-field formulation, the situation becomes more subtle at special points of the parameter space. In particular, at the degenerate point the maximally symmetric vacua coalesce and the FG expansion may contain a nonvanishing $g_{(1)ij}$, thereby introducing an additional holographic source and weakening the standard AdS falloff.

In three-dimensional Einstein-AdS gravity, the standard GHY term supplemented by an intrinsic counterterm is equivalent to a prescription given by one-half of the GHY term~\cite{Miskovic:2006tm}. The difference between both approaches reduces, at the conformal boundary, to the two-dimensional Euler density. The latter is still holographic, as it defines a finite response to the metric at the boundary of AdS spacetimes. This equivalence provides the natural extension of the extrinsic prescription to NMG.

For Brown-Henneaux boundary conditions in NMG, a boundary term linear in the extrinsic curvature is sufficient to renormalize both the action and its variation. This result is not restricted to locally AdS geometries: it applies to nonconstant-curvature configurations with $g_{(1)ij}=0$ as well. The resulting holographic stress tensor has the same form as in Einstein gravity, multiplied by the effective coupling $\varpi_-$. In particular, this explains the vanishing of this quantity and, therefore, of the conserved charges of Einstein solutions at the critical point.

The exact agreement between the holographic stress tensor obtained here and that of Hohm and Tonni~\cite{Hohm:2010jc} at the conformal boundary was found. Their construction writes down the curvature-squared part of the action in terms of an auxiliary field, together with a generalized GHY term and an intrinsic counterterm. Once the auxiliary field takes the value of the Schouten tensor on shell, its FG expansion gives precisely the holographic stress tensor of the extrinsic prescription. Out of the special points of the NMG parameter space, the two different renormalization schemes encode the same holographic data. The current construction provides a direct geometric formulation of existing results without introducing auxiliary variables.

In the degenerate case, that is $\varpi_+=0$, the asymptotic equations no longer constrain $g_{(1)ij}=0$. Although the new mode does not generate an additional divergence in the bulk action, the boundary term needed to renormalize the Brown-Henneaux sector produces an extra divergence proportional to $\tr g_{(1)}$. Finiteness of the action, proper holographic description, and finiteness of the conserved charges can be simultaneously achieved by introducing a quadratic extrinsic term [see Eq.~\eqref{GBdyaction}]. The finite variation contains two stress tensors, associated to two different sources at the conformal boundary. This structure is reminiscent of the appearance of additional boundary data as in four-dimensional Conformal Gravity.

The Noether-Wald formalism provides an independent verification of the construction proposed here. For constant-curvature spaces, the Noether prepotential reduces to the Einstein expression multiplied by $\varpi_-$. Thus, conserved charges for the BTZ black hole agree with the standard NMG result. For nonconstant-curvature AdS waves satisfying Brown-Henneaux boundary conditions (see Ref.~\cite{Ayon-Beato:2009cgh}), the charge associated with their null Killing vector vanishes. Most importantly, at the degenerate point, the same prescription gives, for the rotating hairy black hole, the charges in Eq.~\eqref{chargesOTT}, in exact agreement with the charges previously obtained by independent methods. Thus, the extrinsic counterterms derived here reproduce previous results in both the nondegenerate and degenerate sectors.

The present analysis does not include logarithmic modes arising at the critical point~\cite{Clement:2009ka,Hohm:2010jc,Giribet:2010ed}. Extending the present construction to logarithmic boundary conditions would therefore be a natural next step. More generally, it would be interesting to determine the relation between vacuum degeneracy, holographic sources, and extrinsic renormalization for other higher-curvature theories admitting weakened AdS boundary conditions. These questions define natural directions for future work.

\acknowledgments
The authors thank Giorgos Anastasiou, Gabriel Arenas-Henriquez, Nicolás Cáceres, Luca Ciambelli, Felipe Díaz, Gastón Giribet, Hernán González, Julio Oliva, and Ricardo Troncoso for their valuable comments and discussions. Further insight on the degenerate case was provided by K. Lucero. This work is partially supported by the Agencia Nacional de Investigación y Desarrollo (ANID) through Fondecyt Regular grants No.~1230492, 1231779, 1240043, 1240048, 1240955, 1251523, 1252053, and 1261016. LS acknowledges the partial support from Fondecyt Regular No.~1240043.

\appendix

\section{Generalized Kronecker delta} \label{sec:gen-kdelta}

The generalized Kronecker delta is defined by 
\begin{equation}
    \delta^{\mu_1 \ldots\mu_p}_{\nu_1 \ldots \nu_p} = p!\, \delta^{[\mu_1}_{[\nu_1} \cdots \delta^{\mu_p]}_{\nu_p]} =   \det
    \begin{pmatrix}
        \delta^{\mu_1}_{\nu_1} & \cdots & \delta^{\mu_1}_{\nu_p} \\
        \vdots                 & \ddots & \vdots                 \\
        \delta^{\mu_p}_{\nu_1} & \cdots & \delta^{\mu_p}_{\nu_p}
    \end{pmatrix}\,.
\end{equation}
In $D \geq p \geq k$ dimensions, its contraction satisfies 
\begin{equation}
\delta^{\mu_1\ldots\mu_p}_{\nu_1\ldots\nu_p}\delta^{\nu_1}_{\mu_1}\dots\delta^{\nu_k}_{\mu_k}=\frac{(D-p+k)!}{(D-p)!}\delta^{\nu_{k+1}\ldots\nu_p}_{\mu_{k+1}\ldots\mu_p} \,. \label{footnoteKdelta}
\end{equation}

\section{Gauss-normal foliation} \label{sec:GN-FG}

For the Gauss-normal metric~\eqref{GNcoord} and the extrinsic curvature convention introduced below Eq.~\eqref{GHY}, the nonvanishing Christoffel symbols are 
\begin{subequations} \label{ChristoffelGN}
    \begin{align}
        \Gamma^z_{zz} &= \frac{\partial_zN}{N} \,, &\Gamma^z_{ij}&= \frac{1}{N}K_{ij}\,, \\
        \Gamma^i_{zj} &= -NK^i_j \,, &\Gamma^i_{jk} &= \varGamma^i_{jk}(h) \,.
    \end{align}
\end{subequations}
Here, $\varGamma^i_{jk}(h)$ denotes the Levi-Civita connection of the induced metric $h_{ij}$. The independent components of the bulk Riemann tensor are 
\begin{subequations} \label{RiemannGN}
    \begin{align}
        R^{ij}_{kl} &= \mathcal R^{ij}_{kl} - K^i_k K^j_l + K^i_l K^j_k \,, \\
        R^{zi}_{zj}&=  \frac{1}{N}\partial_z K^i_j - K^i_k K^k_j \,, \\
        R^{zi}_{jk} &=  \frac{2}{N} \mathcal D_{[j} K^i_{k]} \,,
    \end{align}
\end{subequations}
where $\mathcal R^{ij}_{kl}$ and $\mathcal D_i$ are the intrinsic Riemann tensor and the covariant derivative associated with $h_{ij}$. Their contractions give 
\begin{subequations}
    \begin{align}
        R^z_z &= - K^i_j K^j_i + \frac{1}{N}\partial_zK\,, \\ 
        R^z_i&= -\frac{1}{N}\mathcal D_j \left( K^j_i - \delta^j_i K \right) \,, \\
        R^i_j &= \mathcal R^i_j - KK^i_j + \frac{1}{N}\partial_z K^i_j \,,  \\
        R &= \mathcal R -K^2 -K^i_j K^j_i + \frac{2}{N} \partial_z K \,. 
    \end{align}
\end{subequations}

Using the definition of the Schouten tensor in Eq.~\eqref{Schouten}, its Gauss--normal components become 
\begin{subequations} \label{Schouten-comp}
    \begin{align}
        S^i_j &= \mathcal R^i_j - K^i_jK+\frac{1}{N}\partial_z K^i_j \nonumber \\
        &\quad + \frac{1}{4} \delta^i_j \left( K^k_l K^l_k + K^2 - \frac{2}{N}\partial_z K - \mathcal R \right) \,, \\ 
        S^z_z &= -\frac{1}{4} \left( \mathcal R + 3 K^i_j K^j_i - K^2 - \frac{2}{N} \partial_z K   \right) \,, \\ 
        S^z_j &= -\frac{1}{N} \mathcal D_i \left( K^i_j - \delta^i_j K \right)  \,, \\
        S^i_i &= \frac12 \left( \mathcal R - K^2 + K^i_j K^j_i \right)\,.
    \end{align}
\end{subequations}
For the power-law sector of the FG expansion~\eqref{FGexp}, boundary indices are raised and lowered with $g_{(0)ij}$. Substitution of the FG expansion into Eq.~\eqref{Schouten-comp} gives 
\begin{subequations}
    \begin{align}
        S^z_z &= - \frac{1}{2\ell^2} + \frac{z^2}{16 \ell^4} \left[ \tr g_{(1)}^2 + \big( \tr g_{(1)}\big)^2 - 4 \left( 2 \tr g_{(2)} + \ell^2 \mathcal R_{(0)} \right) \right] + \mathcal{O}(z^3)\,, \\ 
        S^i_j &= -\frac{1}{2\ell^2}\delta^i_j
    +\frac{z}{2\ell^3}g_{(1)j}^{i}
    +\frac{z^2}{16\ell^4}
    \Bigg[
        -4g_{(1)j}^{i}\tr g_{(1)} \nonumber \\
        &+\delta^i_j
        \left(
            -3\tr g_{(1)}^2
            +\big(\tr g_{(1)}\big)^2
            +4\left[
                2\tr g_{(2)}
                +\ell^2\mathcal R_{(0)}
            \right]
        \right)
    \Bigg]
    +\mathcal O(z^3)\,.
    \end{align}
\end{subequations}
The corresponding components of the Cotton tensor behave as 
\begin{subequations} \label{CottonFG}
    \begin{align}
         C_{i}^{\,zj}&=\frac{z^{3}}{8\ell^{6}}\Big(2g_{(1)i}^{j}\tr g_{(1)}-2g_{(1)i}^{k}g_{(1)k}^{j} +\delta_{i}^{j}\left(\tr g_{(1)}^{2}-(\tr g_{(1)})^{2}\right)\Big)+\mathcal{O}(z^{4})\,, \\
        C_{i}^{\,jk}	&=\frac{z^{3}}{\ell^{5}}\Big(\delta_{i}^{j}\mathcal{D}_{(0)}^{[l}g_{(1)l}^{k]}+\delta_{i}^{k}\mathcal{D}_{(0)}^{[j}g_{(1)l}^{l]}\Big)+\mathcal{O}(z^{4})\,.
    \end{align}
\end{subequations}

\bibliographystyle{JHEP}
\bibliography{References.bib}

\end{document}